\documentclass[preprint,psfig]{aastex}
\usepackage{mkfig}

\received{}
\accepted{}
\newcommand{\msolar}{\mbox{\,$M_{\odot}$}}
\slugcomment{Submitted to the Astrophysical Journal (Letters), 8 Jan. 2001}

\lefthead{Avillez and Mac Low}
\righthead{Mushroom-Shaped Structures}

\begin{document}

\title{Mushroom-Shaped Structures as Tracers of Buoyant Flow in the
Galactic Disk}  

\author{Miguel A. de Avillez and Mordecai-Mark Mac Low}
\affil{Department of Astrophysics, American Museum of Natural History, \\
Central Park West at 79th Street, New York, NY 10024, U.S.A.;
mavillez, mordecai@amnh.org}
\begin{abstract}
Recent H{\sc i} emission observations of the Southern Galactic hemisphere
have revealed a mushroom-like structure extending from
$z=-70$ to $-450$ pc, composed of a stem and a cap.  Similar
structures occur in three-dimensional simulations of a dynamic
galactic disk driven by isolated and clustered supernovae.  Using
these simulations, we show that hot gas in the Galactic disk that is
not evacuated through chimneys expands into the cooler gas of the
thick disk, forming mushroom-shaped structures. This new class of
objects traces buoyant flow of hot gas into the thick disk.
\end{abstract}
\keywords{Hydrodynamics -- ISM: structure -- Galaxy: evolution --
Galaxy: structure -- Galaxy: general}

\section{Introduction}
H{\sc i} emission maps from the Canadian Galactic Plane Survey (Taylor 1999,
Higgs 1999) reveal a structure resembling a mushroom, with a stem and
a fragmenting cap (English et al.\ 2000).  Assuming a distance of
3.8~kpc, the stem has a length of $\sim230$ pc and the cap has a width
of $\sim175$ pc. The top of the cap lies $\sim450$ pc below the
Galactic plane. This feature covers a total velocity width of $\sim24$
km s$^{-1}$, with the cap redshifted from the stem by $\sim5$ km
s$^{-1}$.  English et al.\ (2000) estimate that the Mushroom has a
kinetic energy of $\sim 2 \times 10^{50}$ ergs and an H{\sc i} mass of at
least $1.55\times 10^{5} \msolar$. The column density contrast between
the Mushroom and the ambient medium is no more than five, with the
Mushroom having N$_{H} \simeq 6\times 10^{20}$ cm$^{-2}$.

Mashchenko et al.\ (1999) and English et al.\ (2000) used thin-shell and
two-dimensional hydrodynamical simulations to study the origin and
dynamics of the Mushroom. They modelled stellar winds and supernovae
(SNe) in a smooth, stratified background atmosphere, and concluded
that the observational data is consistent with a single off-plane SN
having a total energy of at least $\sim10^{51}$ erg s$^{-1}$.  In
their model, the hot gas produced by the single supernova remnant
(SNR) expanded buoyantly into a smooth gas disk.

In this Letter, we use the global models of Avillez (2000) to show
that such mushroom-shaped structures can result from the buoyant rise
of bubbles out of pools of hot disk gas created from both isolated
SNRs and mergers of isolated SNRs.

\section{Supernova-Driven ISM Modelling}

\subsection{Simulations}

Large-scale modelling of the interstellar gas in the disk and halo has
been carried out by Avillez (2000) with a three-dimensional
hydrodynamical code that uses adaptive mesh refinement (AMR).  The
scheme uses the piecewise parabolic method (PPM) of Collela and
Woodward (1984) in combination with the AMR scheme of Berger \&
Colella (1989), combined with the subgridding scheme of Bell et al.\
(1994).

The model includes a fixed gravitational field provided by the stars
in the disk, and an ideal-gas equation of state.  Radiative cooling is
treated assuming collisional ionization equilibrium, using a cooling
function following Kahn (1976) for $10^{5} < T < 5\times 10^{6}$ K,
Dorfi (1997) for $5\times 10^{6} < T < 10^{7}$ K, and a fit to the
curves of Shapiro \& Moore (1976) for the remaining
temperatures. Thermal conduction is not included due to the numerical
complexity and the lack of strong observational evidence for its
strength in the ISM (e.g.\ Mac Low 1999).  Its inclusion may well
change our quantitative results, but we believe that our qualitative
results on the behavior of hot gas in the ISM will continue to hold.

The computational domain contains a section of the Galaxy with an area
of 1 kpc$^{2}$ and vertical extension from -10 to 10 kpc.  The
innermost edge lies 8.5 kpc from the Galactic centre.  The
computational grid has a resolution of 10 pc, except in the layer
between -250 and 250 pc, where three levels of AMR are used, yielding
a finest resolution of 1.25 pc.

Type Ib, Ic, and II~SNe are set up at the beginning of their Sedov
phases, with radii determined by their progenitor masses, at a rate
compatible with observations (Capellaro et al.\ 1997), normalized to
the volume under study. Sixty percent of the SNe are set up within
associations (Cowie et al.\ 1979), and the rest are set up at random
sites.

\subsection{Global Evolution of the ISM}

The disk and halo gas is initially smoothly distributed in hydrostatic
equilibrium with the gravitational field, with scale heights given by
Dickey \& Lockman (1990) and Reynolds (1987).  SNe start occurring at
time zero, and the system is evolved for up to 1 Gyr.  After the first
200 Myr, the system reaches a statistically steady state that shows a
dynamic equilibrium between upward and downward flowing gas.

Once disrupted by the SN explosions, the disk never returns to its
initial state.  Instead, regardless of the initial vertical distribution
of the disk gas, a thin disk of cold gas forms in the Galactic plane,
and, above and below, a thick inhomogeneous gas disk forms.  The code
does not explicitly follow ionization states, but we can trace gas
with temperature $T\leq 10^{4}$ K and scale height of 180~pc, which we
designate H{\sc i}, and gas with $10^{4} \le T\le 10^{5}$ K and scale
height of 1~kpc, which we designate H{\sc ii}. These distributions
reproduce those described in Dickey \& Lockman (1990) and Reynolds
(1987), respectively.  The upper parts of the thick H{\sc ii} disk
form the disk-halo interface, where a large scale fountain is set up
by hot ionized gas escaping in a turbulent convective flow.

The thick gas disk is punctured by chimneys that result from
superbubbles occurring above the plane of the thin disk at
$\left|z\right|\sim 100$ pc.  As they grow, they elongate along the
$z$-direction, owing to the local stratification of the ISM. Chimneys
in the simulation typically have widths of approximately 120 pc. They
inject high temperature gas directly from the Galactic disk into the
halo, breaking through the warm neutral and ionized layers that
compose the thick disk. This hot gas then contributes to the Galactic
fountain.

\section{Buoyant Outflow From the Galactic Plane}

\begin{figure*}
\centering
\includegraphics[angle=-90,width=0.8\hsize]{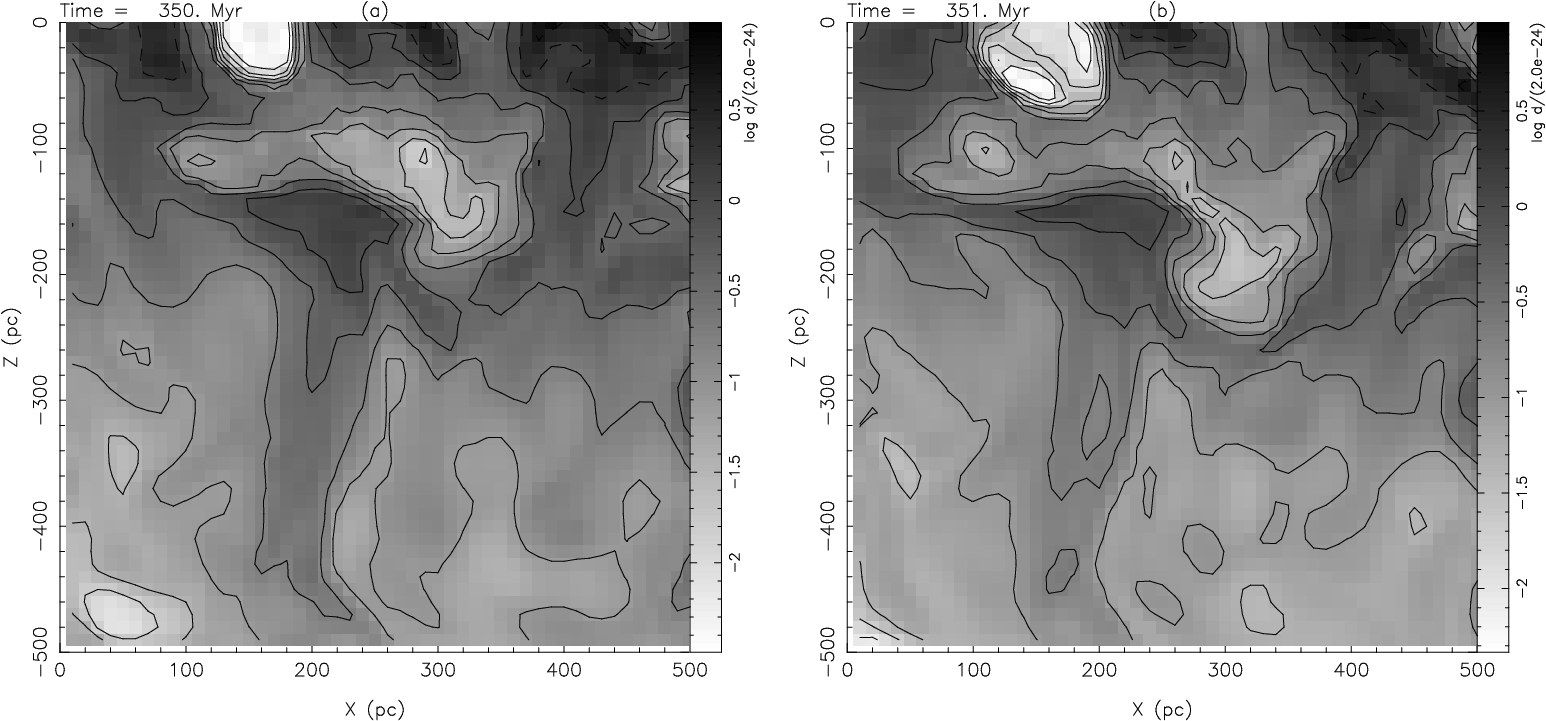}\\
\includegraphics[angle=-90,width=0.8\hsize]{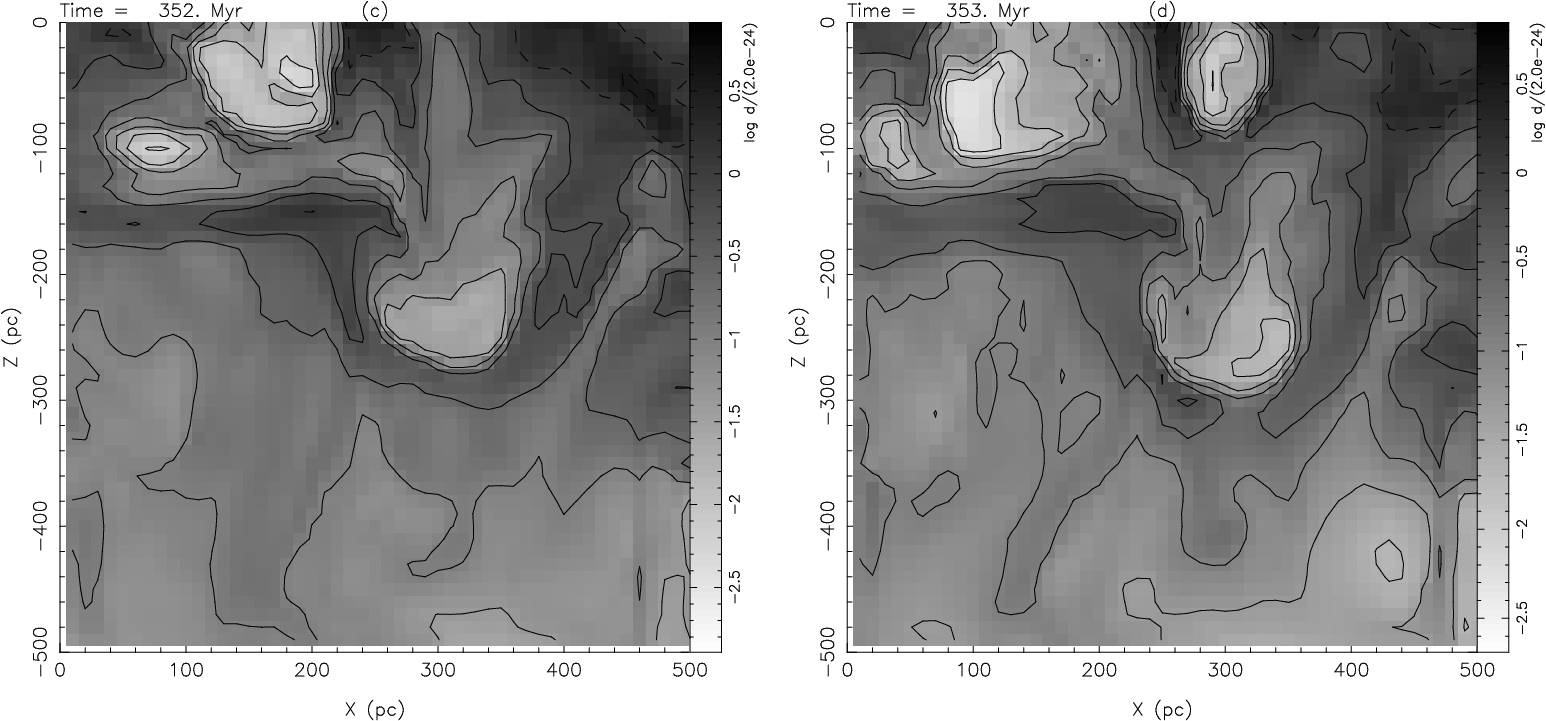}
\caption{Two-dimensional cuts through the three-dimensional density
distribution from the simulations show buoyant expansion of hot,
low-density gas into the thick gas disk. A mushroom-shaped structure
is formed as the ascending gas expands above the 180 pc scale height
of the atmosphere.  Note the narrow stem that forms as the cap rises.
\label{denfig}}
\end{figure*}

About forty percent of the hot gas generated in the disk does not flow
through chimneys. This gas still has too much energy to be held
gravitationally to the stellar disk though, so it expands buoyantly
into the thick gas disk, accelerating through the cooler gas in the
thin disk, with characteristic velocities of some 10 km~s$^{-1}$.  As
the hot gas rises, it entrains surrounding material to form a stem,
often narrow, whose width does not depend on the scale height, as
shown in Figure~\ref{denfig}.  The resulting structure resembles a
mushroom composed of a cap and a stem.  The cap drives the motion of
the mushroom. During the first million years of evolution the mushroom
consists of a rising low density cavity.  The ascending gas comes from
single SNRs as well as from pools of hot gas formed in the disk by
isolated SNRs that merged together. On average, the hot gas produced
by isolated SNe flows upward at a rate integrated over the disk of
$\sim3 \msolar$ per year on each side of the Galactic plane, as
measured at $z=140$ pc (Avillez 2000).

As the mushroom comes to equilibrium with the surrounding
medium and cools down, there is a reversal in the column density
contrast between the mushroom and the surrounding medium, with the
mushroom cap sometimes going from a negative to a positive
contrast. However, the positive column density enhancement is rather
modest, making it very difficult to observationally identify
mushrooms.  

\begin{figure*}
\centering
\includegraphics[angle=-90,width=3.in]{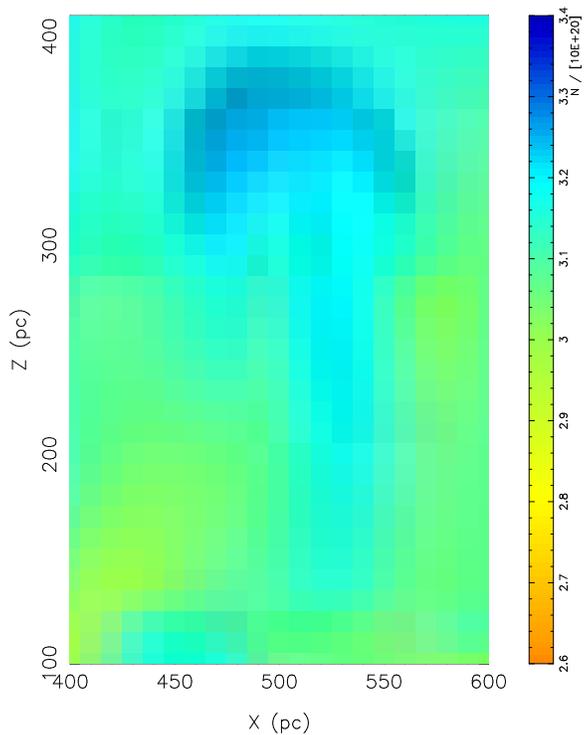}
\caption{Column density map of a mushroom located in the northern
hemisphere after it reached the stabilization point. The color scale
is in units of $N/10^{20}$ cm$^{-2}$. This map was determined by
integrating the column density along the $y-$direction of the
three-dimensional data cube over a pathlength of 1 kpc. The maximum
and minimum values in the scale are 3.4 and 2.6, respectively. There
is a positive column density contrast between the mushroom and
surrounding medium. However, this contrast is very low.
\label{colfig}}
\end{figure*}

\begin{figure}
\centering
\epsscale{0.5}
\includegraphics[angle=-90,width=3.in]{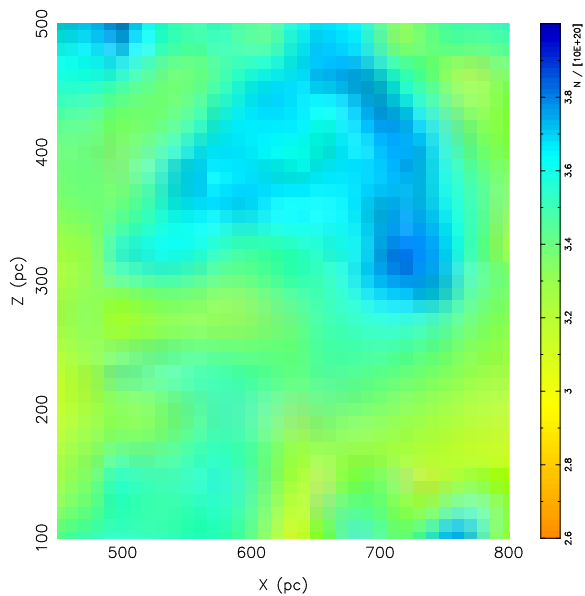}
\caption{Column density map of a mushroom that developed in the
northern hemisphere. The color scale are is in units of $N/10^{20}$
cm$^{-2}$. The mushroom evolved from negative to positive column
density enhancement. The maximum column density in the mushroom is
$4\times 10^{20}$ cm$^{-2}$ (in the cap) while the surrounding medium
has a column density of some $3.2\times 10^{20}$ cm$^{-2}$.
\label{mushroom}}
\end{figure}

Figures~\ref{colfig} and \ref{mushroom} show two mushrooms that
developed at different points in the simulation.  Both structures
cooled to ambient temperatures and have column density enhancements.
On average the ambient column density is $3-3.2 \times 10^{20}$
cm$^{-2}$, while that of the mushrooms is $3.4-4\times 10^{20}$
cm$^{-2}$. The caps have the largest column density enhancement (the
bluish regions inside the caps), while the stems of the two mushrooms
have lower column densities lower than their caps (light bluish
regions in the stems).  The mushroom in Figure~\ref{colfig} was created
by a single SN, while the larger one in Figure~\ref{mushroom} was
created by three SNe.

We can roughly measure the mass ratio $\eta$ of the cap to stem of a
mushroom by estimating the area $A$ and average column density
$\bar{N}$ of each, and taking the mass $M = A\bar{N}$.  The mushroom
shown in Figure~\ref{colfig} has a cap with $A\sim 8000$ pc$^{2}$ and
$\bar{N}=3.3 \times 10^{20}$~cm$^{-2}$; measuring the stem similarly
yields a mass ratio $\eta \sim 2.4$, and a total mass of
$3.7\times10^4 \msolar$.  The mushroom in Figure~\ref{mushroom} has a
mass ratio of 4.4 and a total mass of $1.5\times 10^{5} \msolar$.

The observations of English et al.\ (2000) show the Mushroom to have a
mass of $1.5\times 10^{5} \msolar$ and a mass ratio of three (while a
classical superbubble would have a ratio of $1/3$). These observations
are in rough agreement with the mushroom we show in
Figure~\ref{mushroom}, suggesting that the Mushroom has been created
by 3--4 SNRs.  The greatest column density contrast between the
mushrooms shown in Figures~\ref{colfig} and \ref{mushroom} and the
ambient medium varies between 1.1 and 1.25, while English et al.\
(2000) detected a contrast of roughly two over the velocity range of
the observed structure.

\section{Discussion and Comparison with Observations}

The simulations show that the disk gas is populated with strucures
resembling mushrooms, resulting from hot gas buoyantly convecting
through the thick gas disk.  The dynamics of these structures resemble
the expansion of nuclear fireballs in the Earth's atmosphere.  On the
Earth, as the fireball passes through the tropopause it halts its rise
and spreads because of the turnover in the temperature profile. The
thick gas disk has an isothermal distribution, so there is no
equivalent to the terrestrial tropopause.  However, the cap still
halts when it expands to buoyant equilibrium with the surrounding
atmosphere at a stabilization height $h$.

Let us estimate this height analytically. The simulations produce a
gas distribution similar to that of Dickey \& Lockman (1990), but for
simplicity in our analytical calculation we assume that the thick disk
is isothermal with a temperature of $T_{o}=10^{4}$~K and constant
gravitational acceleration in the $z-$direction.  Thus we approximate
the thick disk with an exponential distribution given by
$\rho_{o}(z)=\rho_{o}(0)\exp(-z/H)$, where $\rho_{o}(0)$ is the
density at the injection level and $H=k_{B} T_{o}/m g$ is the scale
height of the medium. We choose somewhat arbitrarily a value of the
gravitational acceleration inside the stellar disk $g\sim 5\times
10^{-9}$ cm s$^{-2}$ (Kuijken \& Gilmore 1989), giving $H\sim 75$ pc.

The hot gas is lighter than its surroundings, so it rises, with
buoyant force acting on it of $F(z)=g[\rho_{o}(z)-\rho_{i}(z)]$, where
$\rho_{o}(z)$ and $\rho_{i}(z)$ are the densities of the ambient
medium and of the hot interior gas, respectively.  As the hot gas
rises, it maintains pressure equilibrium with the isothermal,
hydrostatic surrounding gas, and continues to rise so long as a
density difference is maintained.

The temperature difference between the hot gas and its surroundings is
expressed in terms of the internal temperature gradient of the hot
gas.  Because the hot gas behaves adiabatically, its internal
temperature gradient is given by
\begin{equation}
\label{eq1}
dT_{i}=\left(\frac{\partial T_{i}}{\partial z}\right)_{ad}dz,
\end{equation}
with
\begin{equation}
\label{eq2}
\left(\frac{\partial T_{i}}{\partial z} \right)_{ad} =
- \left(\frac{\gamma-1}{\gamma}\right)\frac{T_{i}}{P_{i}}.
\end{equation}
where $\gamma=5/3$ is the adiabatic parameter and $T_{i}(z)$ and
$P_{i}(z)$ are the temperature and pressure of the hot gas
at height $z$. Noting that the hot gas remains in pressure equilibrium,
\begin{equation}
\label{eq3}
P_{i}(z)=P_{i}(0)+P_{o}(0)\left(e^{-z/H}-1\right), 
\end{equation}
and integration of equation~(\ref{eq1}), with the help of
equations~(\ref{eq2}) and~(\ref{eq3}), gives
\begin{equation}
 T_{i}(z)=T_{i}(0)e^{-z (\gamma -1) / (\gamma H)}.
\end{equation}
Therefore, the density of the ascending hot gas is
$\rho_{i}(z)=\rho_{i}(0)\exp[-z/ (\gamma H)]$.  In these equations,
$z=0$ refers to the injection level, rather than the plane of the galaxy.

The gas stops rising when the buoyant force vanishes, and 
$\rho_{o}(z)=\rho_{i}(z)$. This allows 
the determination of the stabilization
point, whose location above the injection level is
\begin{equation}
 h=5.75 H\log\left(\frac{\rho_{o}(0)}{\rho_{i}(0)}\right).
\end{equation}
Taking the scale height of the isothermal, exponential atmosphere that
we assumed, $H\sim 75$ pc, hot gas with a density of $\rho_{i}(0)\sim
10^{-26}$ g cm$^{-3}$ will stabilize at some 435 pc above the
injection level.  This is slightly above the value observed in the
Mushroom and found in the simulations, most likely due to radiative
cooling, as we now discuss.

The adiabatic approximation for the buoyant rise only holds if the hot
gas takes longer to radiatively cool than to rise.
Let us assume that hot gas has its origin in the supernova remnant
that generated the Mushroom. As the remnant comes into pressure
equilibrium with the surrounding medium, it contains hot gas that
eventually will mix with the surrounding medium with an
increase in specific energy of
\begin{equation}
E_{o}=\frac{5}{2}\frac{P_{i}(0)}{\rho_{i}(0)} \simeq (2.5\times
10^{14}\mbox{~erg g$^{-1}$}) P_{12} \rho_{26}^{-1},
\end{equation}
where $P_{12} = P_{i}(0)/(10^{-12}$ dyn cm$^{2})$ and $\rho_{26} =
\rho_{i}(0)/(10^{-26}$ g cm$^{-3})$ are the pressure and density of
the hot gas. The adiabatic parameter for the mixed gas is then
\begin{equation}
\kappa  = 
      \left(\frac{2}{5}E_{o}\right)^{5/3}P_{i}(0)^{-2/3}
       =  2.15\times10^{31}  P_{12} \rho_{26}^{-5/3},
\end{equation}
and the time the gas takes to cool is (Kahn 1976)
\begin{equation}
t_{cool} = \frac{\kappa^{2/3}}{q}
         =  (8.3 \mbox{~Myr})  P_{12}^{2/3} \rho_{26}^{-10/9},
\end{equation}
where $q=4\times 10^{32}$ cm$^{6}$~g$^{-1}$~s$^{-4}$ is a constant
dependent on the mean mass of atoms and ions present in the plasma.
This is a time much larger than the time needed for a supernova
remnant to come into pressure equilibrium with the ISM, so the hot gas
must expand upwards into the thick gas disk.

The dynamical time for the hot gas to reach the stabilization point is
$t_{dyn}=(h/g)^{1/2} \simeq 16.4$~Myr for our usual parameters,
roughly twice the cooling time. Thus, cooling does act during
the buoyant rise of the hot gas, reducing the interior pressure and
expansion during the rise, and therefore increasing the interior
density and ultimately reducing the height of the stabilization point
$h$ to that found in the simulations.

The description presented here applies to the mushroom-like structure
described by English et al.\ (2000). We have shown that the structures
found in the simulations resemble the observed structure in shape,
size, vertical extension, and sign of column density enhancement.

In their numerical simulations, Mashchenko et al.\ (1999) and English
et al.\ (2000) attempted to reproduce the Mushroom and the buoyant
rise of the bubble that originated it.  In their toy models they used
artificially low density, low scale-height ($\sim 60$ pc) atmospheres
and an initial SN remnant with a temperature of some $10^{7}$ K. The
simulations described in this Letter show that such a low scale height
is not a requirement, and that structures similar to the Mushroom can
occur in a fully dynamical, supernova-driven disk with a turbulent
density distribution on average reproducing the observed distribution
found by Dickey \& Lockman (1990).  

The temperature of the rising gas in our model is initially only
$10^{6}$ K, and decreases further as a result of its adiabatic
expansion. Thus, no observable X-ray emission is expected by the time
a mushroom is fully developed with a column density higher than the
surrounding medium.

Two issues remain: what determines the column density contrast between
the mushrooms and the surrounding medium; and how frequently does this
new class of objects occur in the Galaxy?  The column density contrast
may result from enhanced cooling due to the compression of the
mushroom cap producing higher densities in the cap. A more physical
treatment of the cooling may lead to the higher observed column
densities.

The number of mushrooms that form is roughly correlated with the
number of isolated SNe distributed in the stellar disk 
Assuming that the distribution of type~Ib+c and II supernovae follow
the distribution of the molecular gas, their rate decreases with the
galactocentric radius (Ferri\`ere 1998), and therefore, the number of
mushrooms will decrease as well at similar rate.  Low density cavities
with small sizes may be detectable as regions with column density
deficits during the first stages of the expansion, but later these
become very hard to detect because of the low column density contrast
between them and the ISM.  The simulations indicate that, even
neglecting Type~Ia SNe, at least $40\%$ of the hot gas in the thin
disk, heated by isolated SNe, should eventually form mushrooms.

\acknowledgements

The authors acknowledge discussions with J. English, S. Basu and
K. Zahnle, as well as a detailed referee's report from J. English. This
work was partially supported by an NSF CAREER fellowship to
M.-M. M. L. (AST 99-85392) and by the AMNH through a Coleman Research
fellowship to M. A. Initial work was supported by ESO/FCT (Portugal)
through contract PESO/P/PRO/1189/97 to M. A..

\clearpage

\end{document}